\documentclass[12pt, fleqn]{article}
\usepackage{latexsym,amsmath,amsfonts,amssymb,amsthm}
\usepackage{graphicx}
\usepackage{indentfirst}
\usepackage{cite}
\usepackage{amsbsy}
\usepackage{epsf}
\usepackage[cp1251]{inputenc}

\newtheorem{theorem}{Theorem}

\newtheorem{corollary}{Corollary}

\newcommand{\ba}{\begin{array}}
\newcommand{\ea}{\end{array}}
\newcommand{\be}{\begin{equation}}
\newcommand{\ee}{\end{equation}}
\newcommand{\p} {\partial}

\begin{document}

\begin{center}
 {\Large \bf Lie symmetry properties \\
 of nonlinear  reaction-diffusion equations \\ with gradient-dependent diffusivity}
 \medskip

 {\bf Roman Cherniha$^{a,b}$\footnote{\small   E-mail: r.m.cherniha@gmail.com}}
 {\bf John R. King$^a$\footnote{\small   E-mail: John.King@nottingham.ac.uk} and  Sergii Kovalenko $^c$\footnote{\small   E-mail: kovss82@gmail.com}}
 \\
{\it
  $^a$~\it School of Mathematical Sciences,
University of Nottingham, \\
University Park, Nottingham, NG7 2RD, UK\\
$^b$~Institute of Mathematics, Ukrainian National Academy
of Sciences,\\
 3, Tereshchenkivs'ka Street, Kyiv 01601, Ukraine\\
$^c$~ Department of Physics, Poltava National Technical University,\\
 24, Pershotravnevyi Prospekt, Poltava 36011, Ukraine
   }\\
 \end{center}

\begin{abstract}  Complete descriptions of the  Lie symmetries of a class   of nonlinear  reaction-diffusion equations  with gradient-dependent diffusivity
in one and two space dimensions are obtained. A  surprisingly
rich  set of Lie symmetry algebras depending on    the form of  diffusivity and source (sink) in the equations  is  derived.
It is established that  there exists   a subclass  in 1-D   space
admitting an  infinite-dimensional Lie algebra of invariance so that it  is linearisable.  A special power-law diffusivity with a fixed exponent, which leads to wider Lie invariance of the equations in question in 2-D   space, is also derived. However,  it is shown that the  diffusion  equation  without a source term (which often arises in applications and is  sometimes called the Perona-Malik equation) possesses no rich variety of Lie symmetries depending on the form of gradient-dependent diffusivity. The results of the Lie symmetry classification for  the reduction to lower  dimensionality,  and a  search for exact solutions of the nonlinear 2-D  equation with power-law diffusivity, also  are  included.

\end{abstract}

\newpage

\section{Introduction }

Nonlinear diffusion equations of the form
\begin{equation}\label{1a}
    u_t = \text{div}\left( D(u) \nabla u \right)
\end{equation}
(hereafter $u = u(t, \bar x)$ is an unknown function, $\bar x=(x_1,... x_n)$, $D$  is an arbitrary nonnegative function of its argument,  $\nabla u \equiv (u_{x_1},..., u_{x_n})$ and  the subscripts
$t$, $x_1$ ...  $ x_n$, denote differentiation with respect to these variables)
have an enormous number of diverse applications and have accordingly been widely studied by Lie symmetry method for a long time  (see e.g.
\cite{ovs} for $n=1$,  \cite{narib1970, dor-svi} for $n=2,3$).

An alternative class with similar status can be written in the form
\begin{equation}\label{0}
    u_t = \text{div}\left( D\left( \Omega \right) \nabla u \right), \quad \Omega = |\nabla u|^2,
\end{equation}
and also  has a significant number of applications (some of
 which we  note below) but has been much less studied from symmetry point of view (excepting the case $n=1$). An observation relating to the similar standing of   (\ref{1a}) and  (\ref{0}) is that the later corresponds to a gradient flow
\[ \frac{\p u}{\p t}=- \frac{\delta\textit{L} }{\delta u} \]
involving  the variational derivative  of
\[ \textit{L} = \frac 12 \int\Psi( |\nabla u|^2)d\bar x, \quad  D( \Omega)= \Psi'(\Omega), \quad  d\bar x\equiv dx_1\dots dx_n, \]
while the former corresponds to
\[ \frac{\p u}{\p t}=-\frac{1}{\Psi'(u)} \frac{\delta\textit{L} }{\delta u} \]
with
\[ \textit{L} = \frac 12 \int |\nabla\Psi(u)|^2d\bar x, \quad  D(u)= \Psi'(u). \]
A third equation  of similar status is
\begin{equation}\label{1b}
    w_t = \Psi'(\Delta w), \quad \Delta \equiv \nabla^2
\end{equation}
 which  is reduced to (\ref{1a}), now  with $ D(u)= \Psi''$ by the non-local substitution $u=\Delta w$: this relationship is a non-local one, so the two may  have different Lie symmetry properties. Taking into account this substitution  and making the calculations,     (\ref{1a}) and  (\ref{1b}) can be written as  the system
 \[ \frac{\p u}{\p t}= \frac{\delta\textit{L} }{\delta w}, \quad \frac{\p w}{\p t}= \frac{\delta\textit{L} }{\delta u}, \]
 where
 \[ \textit{L} =  \int\Psi(u)d\bar x =  \int\Psi(\Delta w)d\bar x. \]
 Finally, we in this case  arrive formally at the equation
 \[ \frac{d\textit{L} }{dt}=-\int  |\nabla \frac{\delta\textit{L} }{\delta u}  |^2d\bar x. \]

In one dimension equations (\ref{1a}), (\ref{0})  and  (\ref{1b})  are non-locally  equivalent and we have the hierarchy
(replacing $u$ by $v$ in the second equation)
\[ u_t=  \left( \Psi'(u)\right)_{xx}, \quad v_t=  \left( \Psi'(v_x)\right)_{x}, \quad  w_t= \Psi'(w_{xx}) \]
with $u=v_x$ and $v=w_x$.
Lie symmetry properties of this hierarchy have been
studied   by  \cite{akhatov-ga-ibr-87, akhatov-ga-ibr-91, bl-k}.
However, when source (sink) terms are included, as we do below, the corresponding reaction-diffusion equations are inequivalent even in one dimension.

It  should be noted that
equation (\ref{0}) for $n=2$  was proposed by  Perona and Malik in   \cite{per-ma-1990} as an edge-enhancement model in image processing
and this represent one of the most prominent applications of (\ref{0}),
 stimulating
a great deal of interest in the image processing community during
the last two decades. It is commonly proposed that the Perona-Malik equation
provides a  valuable  algorithm for image segmentation, noise
removal, edge detection, and image enhancement (see, e.g., \cite{china-2012}).
 In  (\ref{0}),  the unknown function $u = u(t, x_1, x_2)$ describes the time evolution from an  original image, while the diffusion coefficient $D$ (hereafter $D_{\Omega} \neq 0$) is a nonnegative function of
the magnitude of local image gradient $\nabla u \equiv (u_{x_1}, u_{x_2})$.   In \cite{per-ma-1990}, the authors proposed two forms for the diffusion coefficient $D$, namely
\begin{equation}\label{0*}
D( \Omega)= \exp\left(-\frac{\Omega}{D_0}\right), \quad  D( \Omega)= \left(1+\frac{\Omega}{D_0} \right)^{-1}
\end{equation}
(here $D_0$ is a positive constant),
which are the most common in image modelling.
Equation  (\ref{0}) with the power-law  diffusion coefficient
\begin{equation}\label{0a}
D( \Omega)= \Omega^{m},    \ m >0
\end{equation}
has also  been used in modelling turbulent gas  flows: see  \cite{barenb-1952}, for example.

Rigorous  mathematical results  for the Perona-Malik equation have been  obtained  mostly under  the one-dimensional
restriction,  i.e.
\begin{equation}\label{0**}
    u_t = \left( D(u_x)  u_x\right)_x
\end{equation}
  and relevant references are  given in \cite{zhang-2006} (see also the earlier papers  \cite{est-vaz-86, kalashnikov-87}). Equation (\ref{0**}) is  sometimes  called  the nonlinear filtration equation and is applied in modelling non-Newtonian flows of fluids.

Here we concentrate ourselves on the  investigation  of  (\ref{0})  and  its generalisations
 by means of the Lie symmetry method. Because their  Lie symmetry properties essentially depend on the space  dimensionality, we examine separately the cases $n=1$ and $n=2$.
  The Lie symmetries of    (\ref{0**}) were completely described  many years ago \cite{akhatov-ga-ibr-91}.
However, to the best of our knowledge, there  are  no  rigorous  mathematical results concerning the Lie symmetry of   (\ref{0}) in higher dimensions. Moreover,     its  extensions involving   a source (sink)  naturally arise in applications  (see, for example,
   \cite{pascal-1992},  \cite{china-2012}). Thus, one may consider
   the  class of  equations
\begin{equation}\label{0***}
   u_t = \text{div}\left( D\left( \Omega \right) \nabla u \right)+Q(u),
\end{equation}
where $D\not \equiv constant$  and $Q$ are arbitrary smooth functions of their arguments. Equations of the form  (\ref{0***}) also  have an obvious gradient-flow interpretation.

In Section 2, a complete description of Lie symmetries  of equation (\ref{0***})
 is performed in the one-dimensional case.
In Section 3, the same is done
 in two dimensions, the special case   (\ref{0})  being  examined separately. It is worth emphasising at the outset that a number of the symmetries obtained are far from obvious by inspection.
  In Section 4,  radially-symmetric reductions of   equations with the  power-law diffusivity  to ODEs together with examples of exact solutions are presented. Finally, we  discuss the results obtained   in order to compare those with the known results  derived  for more familiar nonlinear reaction-diffusion equations.

\section{Lie symmetry of a class of (1+1)-dimensional    equations  }

In this section,  we  consider
the class of equations
\begin{equation}\label{3-0}
    u_t =  D(u_x)  u_{xx} +Q(u),
\end{equation}
where the smooth function  $D(u_x)$ is an analog of the  diffusivity  in  (\ref{0**}) (obviously   \eqref{3-0} is equivalent to  the (1+1)-dimensional  equation  (\ref{0***}) on redefining $D$ suitably).
 In what follows we assume that  $D(u_x)\not \equiv constant$  because
   \eqref{3-0} with the constant diffusivity is  the standard  reaction-diffusion equation, for which the  Lie symmetry classification  was performed in \cite{dor}.
It is also assumed that  $Q(u)\not \equiv constant$
because  a constant source (sink) $Q(u)=q$ in  \eqref{3-0} is removable by the substitution
 \begin{equation}\label{3-0a}
  t = t,  \  x = x,   \  u  \to u - qt,
  \end{equation}
i.e. \eqref{3-0} with $Q(u)=q$  is  equivalent to \eqref{0**}.

First of all we note that equation \eqref{3-0} with $D(u_x)=u_x^{-2}$ is linearisable by the well-known hodograph transformation.   This linearisation  is a direct consequence of  Lie symmetry properties of  \eqref{3-0} with $D(u_x)=u_x^{-2}$.  In fact, by direct application of the Lie algorithm \cite{ovs, olv} one  proves the following  statement (see a crucial step of  the proof  presented below after Theorem 3).

\begin{theorem}\label{t1}
     The  nonlinear equation
     \begin{equation}\label{3-1}
    u_t =  u_x^{-2} u_{xx} +Q(u),
\end{equation} with an arbitrary smooth function  $Q(u)$ admits an  infinite-dimensional algebra of Lie symmetries generated by the operator
 \begin{equation}\label{3-2}
        X^{\infty} =  w(t,u) \partial_x,
    \end{equation}
    where $ w(t,u)$ is an arbitrary solution of the linear PDE
      \begin{equation}\label{3-3}
    w_t =  w_{uu} -Q(u)w_u.\end{equation}
    The hodograph substitution $x \to u, \ u \to x $ transforms \eqref{3-1} exactly to \eqref{3-3} with $w(t,u)=x(t,u)$.
\end{theorem}

Taking into account the special status of $D(u_x)=u_x^{-2}$ , we  consider further equation \eqref{3-0} with $D(u_x)\not=u_x^{-2}.$
In order  to obtain a complete description of Lie symmetry operators of equations belonging to class  (\ref{3-0}),
we start by finding of the group $\mathcal{E}$ of equivalence transformations as the first step. By definition, each transformation from $\mathcal{E}$ transforms an arbitrary fixed equation of the form (\ref{1}) to another equation  (which can be the same equation) from this class.
Nowadays finding the group $\mathcal{E}$ for a class of PDEs is a standard procedure, which consists in application of a well-known algorithm (see e.g. \cite{akhatov-ga-ibr-91}),   hence we present the result only.

\begin{theorem}\label{t1a}
    The group $\mathcal{E}$  of the class of PDEs \eqref{3-0} consists of the group   $\mathcal{E}_c$ of  {\it continuous} equivalence transformations generated by  transformations
   \begin{equation}\label{3-4}
        \tilde{t} = \alpha t + \delta_0, \ \tilde{x} = \beta x+ \delta_1, \ \tilde{u} = \gamma u + \delta_3, \ \widetilde{D} = \frac{\beta^2}{\alpha} \, D, \ \widetilde{Q} = \frac{\gamma}{\alpha} \, Q,
        \end{equation}
     (here $\alpha>0$, $\beta>0$, $\gamma>0$, $\delta_0 $,  $\delta_1$ and $\delta_3$ are arbitrary real constants  )
     and the set of {\it  discrete}  transformations
    \begin{enumerate}
      \item $x \rightarrow - x$;
      \item $t \rightarrow - t$, $D \rightarrow - D$, $Q \rightarrow - Q$;
      \item $u \rightarrow - u$, $Q \rightarrow - Q$.
    \end{enumerate}
   \end{theorem}

\noindent \emph{\textbf{Remark 1.}} The  discrete  transformation  {\it 2}
   is valid under the assumption that time and diffusivity may be negative, which  means that the results presented below are also valid  for the  associated   non-physical (backward)  equations.

   Obviously, the  discrete  transformations listed above can be incorporated into the formulae \eqref{3-4} by allowing  $\alpha$, $\beta$ and $\gamma$ also to be negative, i.e. only  the restriction $\alpha \beta \gamma \not= 0$  takes place.

    Now we find so called the principal algebra (other terminology used for this algebra is the `trivial algebra'  and the  `kernel of maximal invariance algebras')   of  the nonlinear  PDEs class \eqref{3-0}, i.e. the maximal invariance algebra (MAI) admitted by each equation of the form  \eqref{3-0}.  A direct application of the classical Lie algorithm under the assumption that the coefficients $D(u_x)$ and $Q(u)$ are arbitrary smooth functions leads the two-dimensional Abelian    algebra $ALie_2$ generated by the basic operators
     \begin{equation}\label{3-4a}
      X_1 = \partial_t, \ X_2 = \partial_{x}.
     \end{equation}

 In order to obtain a complete description of Lie symmetries  of  the PDE class \eqref{3-0}, one needs to      find all possible pairs $(D, Q)$, for which  the relevant  nonlinear PDE
is invariant under a MAI of a higher dimensionality than two.

\begin{theorem}\label{t1b}
All possible MAIs (up to  equivalent transformations)
of equation (\ref{3-0}) for any fixed
  pair of the functions  $(D, Q)$  are  presented in Table 1.  Any other equation of the form (\ref{3-0}) invariant under a non-trivial Lie algebra  is reduced
  to one of those given in Table 1 by an equivalence transformation from $\mathcal{E}$.
\end{theorem}

\begin{table}
\caption{Lie symmetries  of the class of  equations  \eqref{3-0} with  $D \neq u_x^{-2}$}
\renewcommand{\arraystretch}{1.5}
\tabcolsep=8pt
\begin{center}
\footnotesize{
\begin{tabular}{ccccl}

\hline\hline

 & $D(u_x)$ & $Q(u)$ & Constraints &  MAI  \\

\hline

1. & $\forall$ & $u^{-1}$ &  & $ALie_2, \ X_3 = 2t\partial_t + x\partial_{x}  + u\partial_u$ \\

2. & $\forall$ & $u$ &  & $ALie_2, \ X_3 = e^t \partial_u$ \\

3. & $u_x^k$ & $\varepsilon_1e^{-u}
$ &  & $ALie_2, \  X_3 =(k+2)t \partial_t +  x\partial_x + (k+2)   \partial_u $ \\

4. & $u_x^k$ & $\varepsilon_1 u^m$ & $m \neq  1, 2$ & $ALie_2, \  X_3 = (1-m)t \partial_t + \frac{k+1-m}{k+2} x \partial_{x}+ u \partial_u $ \\

5. & $u_x^k$ & $\varepsilon_1 u^{k+1} + \varepsilon_2 u$ & $k \neq \pm1$ & $ALie_2, \  X_3 = e^{-k\varepsilon_2t} (\partial_t + \varepsilon_2 u \partial_u)$ \\
& & & & \\
6.& $\frac{1}{u_x+\gamma}$ & $
u$ & $\gamma \not=0$ &  $ALie_2, \ X_3 = e^{\varepsilon_1 t} \partial_u$, $X_4 = e^{\varepsilon_1 t} (\partial_t + \varepsilon_1( u+\gamma x) \partial_u)$\\

7. & $u_x$ & $\varepsilon_1 u^2$ & & $ALie_2, \ X_3 = t \partial_t - u \partial_u$, $X_4 = t^2 \partial_t - (2tu + \varepsilon_1) \partial_u$ \\

8. & $u_x$ & $\varepsilon_1 u^2 + \varepsilon_2$ &$\varepsilon_1 \varepsilon_2 = -1$  & $ALie_2, \ X_3 = e^{-2t} (\partial_t + 2(u-\varepsilon_1) \partial_u),$ \\
& & & & $X_4 = e^{2t} (\partial_t - 2(u+\varepsilon_1) \partial_u)$ \\

9. & $u_x$ & $\varepsilon_1 u^2 + \varepsilon_2$ & $\varepsilon_1 \varepsilon_2 = 1$ & $ALie_2, \ X_3 = \cos(2t) \partial_t + 2(\sin(2t) u + \varepsilon_1 \cos(2t)) \partial_u,$ \\
& & & & $X_4 = \sin(2t) \partial_t - 2(\cos(2t) u - \varepsilon_1 \sin(2t)) \partial_u$ \\
& & & & \\
10. & $u_x^k$ & $\varepsilon_2 u$ &  & $ALie_2, \ X_3 = x \partial_{x}  + \left( 1 + \frac{2}{k} \right) u \partial_u$, \\
& & & & $ X_4 = e^{-k \varepsilon_2 t} \, (\partial_t + \varepsilon_2 u \partial_u)$, $X_5 = e^{\varepsilon_2 t} \partial_u$ \\

\hline\hline

\end{tabular}}
\end{center}
\end{table}

\noindent \emph{\textbf{Remark 2.}} In Table 1, by rescalling and/or reflecting (see Theorem 2)  the  coefficients  $\varepsilon_1=\pm 1$ and $\varepsilon_2 = \pm 1$  without loss of generality.  Of course, the relevant function $Q$ in real world applications may contain arbitrarily given non-zero coefficients $\lambda_1$ and $\lambda_2$ instead of these.
This remark is valid also for Table 2 (see Section 3).

\textbf{The proof.} Here we apply the  Lie-Ovsiannikov  approach (the name of  Ovsiannikov arises because he  has published a remarkable paper in this direction, \cite{ovs-1959}) of the Lie symmetry classification, which  is based on the classical Lie scheme
and a set of equivalence transformations of  the  differential equation in question. It should be noted that there are some recent approaches,  which  are  important for obtaining the  so called canonical list of inequivalent equations admitting non-trivial Lie symmetry algebras and allow the solution of  the problem of  the Lie symmetry classification  in a more efficient way then a  formal application of the Lie-Ovsiannikov approach  (see some remarks at the end of this section).

Thus, in order to find Lie  symmetry   operators, one needs
to consider   each  equation  of the form   (\ref{3-0}) (clearly  this should be viewed  a class of PDEs because the functions $D$ and $Q$ are arbitrary smooth functions) as the manifold
 \be\label{2-1}
 {\cal{M}} \equiv  \{  u_t -  D(u_x)  u_{xx} -Q(u) =0 \}   \ee
  in the prolonged
space of the  variables: $t, x, u,  u_t,  u_{x},
u_{xx}, u_{xt},  u_{tt} $.  According to the
definition,  (\ref{3-0})  is invariant under the transformations
generated by the infinitesimal operator \be\label{2-2} X = \xi^0 (t,
x, u)\p_{t} + \xi^1 (t, x, u)\p_{x} +
 \eta(t, x, u)\p_{u},  \ee
if the following invariance condition is satisfied:
\be\label{2-3}
\mbox{\raisebox{-1.1ex}{$\stackrel{\displaystyle
X}{\scriptstyle 2}$}}\big ( u_t -  D(u_x)  u_{xx} -Q(u) \big)
\Big\vert_{\cal{M}} =0. \ee
The operator $ \mbox{\raisebox{-1.1ex}{$\stackrel{\displaystyle  
X
}{\scriptstyle 2}$}} $  
is the second   prolongation of the operator $X$, i.e.  
\[
\mbox{\raisebox{-1.1ex}{$\stackrel{\displaystyle  
X}{\scriptstyle 2}$}}  
 = X + \rho_t\frac{\partial}{\partial u_{t}}+
\rho_x\frac{\partial}{\partial u_{x}}+ 
+\sigma_{xx}\frac{\partial}{\partial u_{xx}}  
+\sigma_{tt}\frac{\partial}{\partial  u_{tt}}+\sigma_{tx}\frac{\partial}{\partial  u_{tx}},
\]
where the coefficients $\rho$ and $\sigma$ with relevant subscripts  
are expressed  via the functions $\xi^0, \xi^1$ and $\eta$
by well-known formulae (see, e.g., \cite{ovs, olv}).

Now we present the so called   system of  determining equations (DEs), obtained by direct  calculations using  the invariance condition  (\ref{2-3})
\be\label{2-4} \Big ( \eta_{x} +(\eta_u-\xi^1_x)u_x- \xi^1_u u_x^2\Big )D'+
 \Big (\xi^0_t- 2\xi^1_x- 2\xi^1_u u_x\Big )D=0, \ee

\[-\eta_t +(\xi^0_t-\eta_u)Q +\eta \dot Q +(\xi^1_t +\xi^1_u Q)u_x +\]
\be\label{2-5}
\Big ( \eta_{xx} +(2\eta_{xu}-\xi^1_{xx})u_x+ (\eta_{uu}- 2\xi^1_{xu})u^2_x - \xi^1_{uu} u_x^3\Big )D
=0 \ee
(hereafter  $D'=\frac{\partial D}{\partial u_{x}}$  and $\dot Q =\frac{\partial Q}{\partial u}$)
together with the standard restriction  $\xi^0= \xi^0(t)$,  which is common  for any evolution second-order PDE.

In contrast to the system of  DEs for the standard RD equation (see those in \cite{dor}), which immediately
specifies  the  structure  of the   unknown functions  $\xi^0, \xi^1$ and $\eta$,  DEs (\ref{2-4}) and (\ref{2-5})
are much more complicated because the diffusivity $D$ depends on $u_x$ (other than on  $u$).
The crucial step is to solve  (\ref{2-4}) w.r.t. $D$, taking into account that  the functions $\xi^0, \xi^1$ and $\eta$ do not depend on $u_x$, so that one may formally write (\ref{2-4}) as follows
\be\label{2-6}  ( e_{0} +e_1u_x- e_2 u_x^2)D'+
 (e_3- 2e_2 u_x)D=0, \ee
 where  $e_k, \  k=0,\dots,3$  correspond to the relevant coefficients in  (\ref{2-4}).

 It has been noted  that there exists a unique case $D=C u_x^{-2}$ (the constant $C$ can be removed by  scaling, see Theorem 2)  when the parameter  $e_2$
 can be arbitrary in (\ref{2-6}) because only in this case does the relationship  $e_2 u_x^2D'=2e_2 u_xD$ apply.
 Thus, (\ref{2-6}) with $D=C u_x^{-2}$ is equivalent to $e_{0}=0, \  e_3=2e_1$, i.e.
 \[   \eta_{x}=0, \quad   \xi^0_t=2\eta_u \]
 while $\xi^1$ is still an arbitrary function. The remaining  DE (\ref{2-5}) with $D=C u_x^{-2}$
 and an arbitrary smooth function $Q$
 has the solution $\xi^1=w(t,u)$, where $w(t,u)$ is the general  solution of the linear equation (\ref{3-3}).
 This means that MAI of  (\ref{3-1}) is infinite-dimensional (see operator (\ref{3-2}))
 As a  result,  (\ref{3-1}) can be linearised (see Theorem 1).  Thus, we assume $D\not=C u_x^{-2}$ in what follows.

 Solving the linear ODE  (\ref{2-6}), one arrives on five essentially different cases depending on the parameters.

{ \it (i)} If $e_2=e_1=e_0=0$ then immediately $e_3=0$ and $D$ is an arbitrary non-constant function.

{\it(ii)}  If  $e_2=0$ and $e_1\not=0$  then the general solutions is
\be\label{2-6ii} D=C \Big(u_x+ \frac{e_0}{e_1}\Big)^{-\frac{e_3}{e_1}}, \quad e_3\not=0. \ee

{ \it(iii)} If  $e_2=e_1=0$ and $e_0e_3\not=0$  then the general solution is
\[ D=C e^{-\frac{e_3}{e_0}u_x}. \]

{\it (iv)} If  $e_2\not=0$ and $e_1^2+e_0^2
   \not=0$  then the general solutions is
   \[ D= C\exp\Big(- \int \frac{ 2e_2u_x-e_3}{e_2u_x^2-e_1u_x-e_0}du_x \Big). \]

{ \it (v)} If  $e_2\not =0$,  $e_1=e_0=0$  and $e_3\not=0$ then the general solutions is
\[ D=C u_x^{-2}\exp\Big(-\frac{e_3}{e_0}u_x^{-1}\Big). \]

Consider  case { \it (i)}. Having  $e_3=e_2=e_1=e_0=0$  and substituting instead of $e_k, \  k=0,\dots,3$,   the relevant coefficients from   (\ref{2-4}), one arrives at a linear system of PDEs with the general solution
\be\label{2-7}     \xi^1=\frac12 \xi^0_t x+b(t), \quad   \xi^1=\frac12 \xi^0_t u+p(t)
\ee
where $\xi^0(t), b(t)$ and $p(t)$ are  arbitrary functions. Substituting  (\ref{2-7}) into   DE (\ref{2-5}),
one immediately derives  a  linear ODE w.r.t. $Q(u)$ because the second line of  (\ref{2-5}) simply vanishes.
If $Q(u)$ is an arbitrary function then only  the principal algebra  (\ref{3-4a})  occurs.
Non-trivial results are obtained only for   $Q(u)=\lambda u^{-1}$ and $Q(u)=\lambda u$. In both cases,  the arbitrary non-zero constant $\lambda$  can be removed from the relevant  equations using the equivalence transformations from Theorem 2. Thus, the cases 1 and 2 in Table 1 have been obtained.

Consider  case { \it (ii)}, which is the most complicated. Because the function $D$  depend only on $u_x$,
one derives from  (\ref{2-6ii})  the equalities $e_3=-ke_1\not=0, \ e_0=\gamma e_1$  and $e_2=0$ (hereafter $k\not=0$ and $\gamma$ are arbitrary constants).
Taking into account the relevant coefficients from   (\ref{2-4}), one obtains the linear system of PDEs
\be\label{2-8}
\xi^0_t- 2\xi^1_x=-k(\eta_u-\xi^1_x), \quad     \eta_x=\gamma (\eta_u-\xi^1_x), \quad \xi^1_u=0.
\ee
There are two essentially different subcases { \it (iia)} $\gamma\not=0$ and { \it (iib)} $\gamma=0$.

In  subcase { \it (iia)}, equations  (\ref{2-8}) immediately produce $\xi^1_{xx}=0$ provided $k\not=-2$
(this special exponent was checked separately; nothing interesting was found, however). Having $\xi^1_{xx}=0$ and equations (\ref{2-8}), one notes that DE (\ref{2-5})  again reduces to a linear ODE w.r.t. $Q(u)$. Moreover, only the  functions  $Q(u)=\lambda u^{-1}$ and $Q(u)=\lambda u$ lead to non-trivial Lie symmetry algebras while the diffusivity is $D=C (u_x+ \gamma)^{k}$. However, these algebras are nothing else but those arising in case 1 and 2 of Table 1 provided $k\not=-1$. The special exponent $k=-1$  was examined and the case 6 of Table 1 was   found
(the constants $C$ and $\lambda$  are removable using  the equivalence transformations from Theorem 2).

In   subcase { \it (iib)} $\gamma=0$,
equations  (\ref{2-8})  take form
\be\label{2-9}
\xi^0_t=(2+k)\xi^1_x -k\eta_u, \quad     \eta_x=0, \quad \xi^1_u=0,
\ee
so that $\xi^1_{xx}=0$ ( $k\not=-2$ otherwise $D=C u_x^{-2}$). Thus,  the remaining  DE (\ref{2-5})  again reduces to a linear ODE w.r.t. $Q(u)$. Its solution  for arbitrary $k\not=-2$ leads to the cases 3--5 and 10 of Table 1.  Moreover,  it was found that  case 5  is  not valid  provided  $k=\pm1$ (see this restriction in Table 1), because
  that for $k=-1$ is simply a subcase of case 4, while   that for $k=1$  leads to the very unusual Lie symmetry algebras presented in the cases 7--9 of Table 1.

  Finally, it was proved by the examination of cases  { \it (iii)}--{\it (v)}  that  there are not any further  equations of the form (\ref{3-0}), which are invariant w.r.t. a MAI of  dimensionality three or  higher.

  The proof is now completed. \hfill $\blacksquare$

\medskip

Thus, we have proved  that there are  11 distinct cases  in which
equation  (\ref{3-0})
admits a non-trivial MAI, i.e. is invariant under three-  or  higher-dimensional  Lie algebra.   The case of  the diffusivity $D(u_x)=u_x^{-2}$ is exceptional because the  MAI is  infinite-dimensional, hence, the corresponding equation is linearisable independently on the form of the function  $Q(u)$. The other 10 cases are presented in Table 1.
It is noteworthy that this classification is surprisingly rich: several of these Lie symmetries  (see, e.g. cases 6--9)  would be unlikely to be identified by  using the known results for   the standard reaction-diffusion equation \cite{dor}.

Finally in this context, we present the following observation. The substitution
\begin{equation}\label{3-0b}
  \tau =  \frac{1}{\varepsilon_2 k}e^{\varepsilon_2 kt},  \  x = x,   \  w=e^{-\varepsilon_2 t}u
  \end{equation}
  reduces the equations listed in cases 5 and 10 to  the equation with $m=k+1$  listed in case 4 and to the equation
  $ w_{\tau} =  w_x^{k} w_{xx} $, respectively.  Thus, Table 1 can be shortened to 8 cases.  However, substitution (\ref{3-0b}) does not belong to the group $\mathcal{E}$, hence  we have found so called  form-preserving transformations \cite{kingston-98} (it should be noted that  this kind of transformations  was introduced earlier  in \cite{kingston-91} but without above terminology)  allowing a reduction of  the number of cases obtained via the classical Lie-Ovsiannikov algorithm (see   extensive discussions in  \cite{ch-king2}, \cite{ch-king4}, \cite{ch-se-ra-08} on this matter).

\section{Lie symmetry of a class of (1+2)-dimensional    equations }

Here we consider a class of    equations  of the form
\begin{equation}\label{1}
    u_t = \text{div}\left( D\left( \Omega \right) \nabla u \right) + Q(u), \quad \Omega = |\nabla u|^2,
\end{equation}
where $u = u(t, x_1, x_2) \in \mathbb{R}$, $t \in \mathbb{R}$, $(x_1, x_2) \in \mathbb{R}^2$; $D(\Omega)$ and $Q(u)$ are arbitrary smooth functions of their arguments.
 It can be easily noted that  equation  (\ref{1}) with  $Q(u)=q \equiv constant $  again reduces to  (\ref{0}), by the substitution
 \begin{equation}\label{2}
  t = t,  \  x_1 = x_1, \  x_2 = x_2, \  u  \to u - qt,
  \end{equation}
  so that we  assume in what follows that either  $Q(u)$ is non-constant or $Q(u)=0$.

\subsection{The case of a non-constant source}

We start from the general case when  $Q(u)$ is an arbitrary  non-constant function. Similarly to the (1+1)-dimensional case, we  find   the group $\mathcal{E}$ of  equivalence transformations as the first step.

\begin{theorem}\label{t2}
    The group $\mathcal{E}_c$ of  {\it continuous} equivalence transformations of the class of PDEs \eqref{1} consists of the  transformations
    \begin{equation}\label{3}
        \tilde{t} = \alpha t + \delta_0, \ \tilde{x}_a = \beta \gamma_{ab} x_b + \delta_a, \ \tilde{u} = \gamma u + \delta_3, \ \widetilde{D} = \frac{\beta^2}{\alpha} \, D, \ \widetilde{Q} = \frac{\gamma}{\alpha} \, Q,
        \end{equation}
    where $\alpha$, $\beta$, $\gamma$, $\delta_k \, (k = 0, \ldots, 3)$, $\gamma_{ab} \, (a, b = 1, 2)$ are arbitrary real constants  satisfying  the conditions  $\alpha > 0$, $\beta > 0$, $\gamma > 0$, and the matrix $(\gamma_{ab}) \in SO(2)$.
\end{theorem}

It can be noted that the set of discrete transformations
    \begin{enumerate}
      \item $x_1 \rightarrow - x_1$, $x_2 \rightarrow - x_2$;
      \item $t \rightarrow - t$, $D \rightarrow - D$, $Q \rightarrow - Q$;
      \item $u \rightarrow - u$, $Q \rightarrow - Q$.
    \end{enumerate}
    also belongs to  the group $\mathcal{E}$, hence, one may claim that
the group of equivalence transformations of the nonlinear  PDEs class \eqref{1} consists of the   transformations defined  by formulas \eqref{3} with arbitrary real non-zero  parameters $\alpha $, $\beta$, and $\gamma $.

As the second  step, we find  the principal algebra
  i.e. the MAI admitted by each equation of the form  \eqref{1}, which comprises the  operators of  translations and rotations.

\begin{theorem}\label{t4}
    The principal algebra  of the PDEs class \eqref{1} is the four-dimensional Lie algebra $ALie_4$ generated by the basic operators
    \[
        X_1 = \partial_t, \ X_2 = \partial_{x_1}, \ X_3 = \partial_{x_2}, \ X_4 = x_2 \partial_{x_1} - x_1 \partial_{x_2}.
    \]
\end{theorem}

The proof of this theorem consists in the direct application of the classical Lie algorithm under the  assumption that the coefficients $D(\Omega)$ and $Q(u)$ are arbitrary smooth functions. The relevant calculations are omitted here because they are simple but cumbersome and the result (namely the operators of  translations and rotations)  is to be expected.

The third  and final  step is  highly nontrivial: to find all pairs $(D, Q)$, when the relevant  nonlinear PDE \eqref{1}
is invariant under a MAI of a higher dimensionality than four. Note that we present  the result obtained in two tables because the diffusion coefficient  $D = \Omega^{-1}$    is special and  this case, which corresponds to a limit case  of the second formula in (\ref{0*}), is especially noteworthy.

\begin{theorem}\label{t5}
All possible MAIs (up to the equivalence transformations)  of equation (\ref{1}) for any fixed
  pair of the function  $(D, Q)$ are presented in Tables 2 and 3, where the  designations $\varepsilon_1=\pm 1$, $\varepsilon_2 = \pm 1$  are used  and  $k \neq 0$, $\lambda \neq 0$, $m$ are arbitrary real constants.
   Any other equation of the form (\ref{1})  invariant under a non-trivial Lie algebra is reduced to one of those given in Tables 2 and 3 by an equivalence transformation from the group $\mathcal{E}$.
\end{theorem}

\begin{table}
\caption{Lie symmetries  of the class of  equations
  \eqref{1}: $D \neq \Omega^{-1}$}
\renewcommand{\arraystretch}{1.5}
\tabcolsep=8pt
\begin{center}
\footnotesize{
\begin{tabular}{ccccl}

\hline\hline

 & $D(\Omega)$ & $Q(u)$ & Constraints &  MAI \\

\hline

1. & $\forall$ & $u^{-1}$ &  & $ALie_4, \ X_5 = 2t\partial_t + x_1\partial_{x_1} + x_2\partial_{x_2} + u\partial_u$ \\

2. & $\forall$ & $u$ &  & $ALie_4, \ X_5 = e^t \partial_u$ \\

3. & $\Omega^k$ & $\varepsilon_1 e^{-u}$ &  & $ALie_4, \ X_5 =2(k+1)t \partial_t+ x_1 \partial_{x_1} + x_2 \partial_{x_2}+$  \\
& & & & $ 2(k+1)    \partial_u $ \\

4. & $\Omega^k$ & $\varepsilon_1 u^m$ & $m \neq 0, 1, 2$ & $ALie_4, \  X_5 = 2t \partial_t + \frac{m-2k-1}{m-2k-1+k(m+1)}\times$ \\
& & & & \quad \quad \ $ (x_1 \partial_{x_1} + x_2 \partial_{x_2}) - \frac{2(k+1)}{m-2k-1+k(m+1)} \, u \partial_u$ \\

5. & $\Omega^k$ & $\varepsilon_1 u^{2k+1} + \varepsilon_2 u$ & $k \neq -\frac{1}{2}, \frac{1}{2}$ & $ALie_4, \ X_5 = e^{-2k\varepsilon_2t} (\partial_t + \varepsilon_2 u \partial_u)$ \\


6. & $\Omega^{\frac{1}{2}}$ & $\varepsilon_1 u^2$ &  & $ALie_4, \ X_5 = t \partial_t - u \partial_u$, $X_6 = t^2 \partial_t - (2tu + \varepsilon_1) \partial_u$ \\

7. & $\Omega^{\frac{1}{2}}$ & $\varepsilon_1 u^2 + \varepsilon_2$ & $\varepsilon_1 \varepsilon_2 = -1$ & $ALie_4, \ X_5 = e^{-2t} (\partial_t + 2(u-\varepsilon_1) \partial_u),$ \\
& & & & $X_6 = e^{2t} (\partial_t - 2(u+\varepsilon_1) \partial_u)$ \\

8. & $\Omega^{\frac{1}{2}}$ & $\varepsilon_1 u^2 + \varepsilon_2$ & $\varepsilon_1 \varepsilon_2 = 1$ & $ALie_4, \ X_5 = \cos(2t) \partial_t + 2(\sin(2t) u + \varepsilon_1 \cos(2t)) \partial_u,$ \\
& & & & $X_6 = \sin(2t) \partial_t - 2(\cos(2t) u - \varepsilon_1 \sin(2t)) \partial_u$ \\

9. & $\Omega^k$ & $\varepsilon_2 u$ &  & $ALie_4, \ X_5 = x_1 \partial_{x_1} + x_2 \partial_{x_2} + \left( 1 + \frac{1}{k} \right) u \partial_u$, \\
& & & & $X_6 = e^{-2k \varepsilon_2 t} \, (\partial_t + \varepsilon_2 u \partial_u)$, $X_7 = e^{\varepsilon_2 t} \partial_u$ \\
\hline\hline

\end{tabular}}
\end{center}
\end{table}

\begin{table}
\caption{Lie symmetries  of the class of  equations \eqref{1}: $D = \Omega^{-1}$}
\renewcommand{\arraystretch}{1.5}
\tabcolsep=8pt
\begin{center}
\footnotesize{
\begin{tabular}{cccl}

\hline\hline

 & $Q(u)$ &  MAI \\

\hline

1. & $\forall$ &  $ALie_4, \ X_5 = x_1\partial_{x_1} + x_2\partial_{x_2}$ \\

2. & $\lambda u^{-1} + \varepsilon_2 u$ &  $ALie_4, \ X_5 = x_1\partial_{x_1} + x_2\partial_{x_2}$, $X_6 = e^{2 \varepsilon_2 t} (\partial_t + \varepsilon_2 u \partial_u)$ \\

3. & $\lambda u^{-1}$ &  $ALie_4, \ X_5 = x_1\partial_{x_1} + x_2\partial_{x_2}$, $X_6 = 2t \partial_t + u \partial_u$ \\

4. & $\lambda u$ &  $ALie_4, \ X_5 = x_1\partial_{x_1} + x_2\partial_{x_2}$ \\
& & $X_6 = e^{2 \lambda t} (\partial_t + \lambda u \partial_u)$, $X_7 = e^{\lambda t} \partial_u$ \\

\hline\hline

\end{tabular}}
\end{center}
\end{table}

\noindent \emph{\textbf{Remark 3.}} Similarly to the (1+1)-dimensional case, the substitution
\begin{equation}\label{6}
  \tau =  \frac{1}{2\varepsilon_2 k}e^{2\varepsilon_2 kt},  \  x_1 = x_1,    \  x_2 = x_2, \  w=e^{-\varepsilon_2 t}u
  \end{equation}
  reduces the equations listed in cases 5 and 9 to  the equation with $m=2k+1$  listed in case 4 and to the equation
  $ w_{\tau} =   \text{div}\left(  |\nabla w|^{2k} \nabla w \right) $, respectively.

  \noindent \emph{\textbf{Remark 4.}} Substitution  \eqref{6} with $k=-1$ reduces the equations listed in cases 2 and 4 of Table 3  to  the equation  listed in case 3 and to the equation
  $ w_{\tau} =   \text{div}\left(  |\nabla w|^{-2} \nabla w \right) $, respectively. Thus, Table 3 consists of two essentially different  cases only; however, both cases have no analogs in Table 2.

\subsection{Nonlinear diffusion equation  with gradient-dependent diffusivity
}

It is quite natural to present the result   for the nonlinear  equation \eqref{0} (hereafter with $n=2$) separately because of  its  importance for direct applications.

\begin{theorem}\label{t8}
    The group $\mathcal{E}$ of equivalence transformations of  the  class of equations   \eqref{0} consists of the group of  continuous equivalence transformations
      \begin{equation}\label{3*}
        \tilde{t} = \alpha t + \delta_0, \ \tilde{x}_a = \beta \gamma_{ab} x_b + \delta_a, \ \tilde{u} = \gamma u + \delta_3, \ \widetilde{D} = \frac{\beta^2}{\alpha} \, D,
      \end{equation}
  (the restrictions on the  parameters are the same as in  \eqref{3})
 and the discrete transformations
    \begin{enumerate}
      \item $x_1 \rightarrow - x_1$;
      \item $t \rightarrow - t$, $D \rightarrow - D$;
      \item $u \rightarrow - u$.
    \end{enumerate}
\end{theorem}

Similarly to  the general class \eqref{1},  the group of equivalence transformations of  \eqref{0} is  represented only by the formulae \eqref{3*},
though  with arbitrary non-zero  parameters $\alpha $, $\beta$, and $\gamma $.

In
 comparison to the principal algebra  of  the   PDE class \eqref{1}, that  of  \eqref{0} is  more complicated but still can be easily calculated by direct application of the Lie algorithm, hence the following statement has been  proved.

\begin{theorem}\label{t10}
     The principal algebra  of   \eqref{0}   is the six-dimensional Lie algebra $ALie_6$ generated by the basic operators
      \begin{equation}\label{4}
        X_1, \ X_2,  \ X_3, \  X_4,
       \ X_5 = \partial_u,  \  X_6 = 2t \partial_t + x_1 \partial_{x_1} + x_2 \partial_{x_2} + u \partial_u
     \end{equation}
     (see $X_1, \ X_2,  \ X_3$, and $ X_4$ in Theorem \ref{t4}).
\end{theorem}

\noindent \emph{\textbf{Remark 5.}} The Lie algebra $ALie_6$ is a semidirect sum of the extended Euclidean algebra with the basic operators $X_2, X_3, X_4, X_6$  and the Abelian algebra with the basic operators $X_1,  X_5$.

 As the final step, we perform  an analysis in order to find all possible functions $D(\Omega)$ leading to extensions of the Lie algebra $ALie_6$.   It turns out that the result is (perhaps surprisingly) rather simple.

\begin{theorem}\label{t11}
     Equation \eqref{0} admits a  MAI of  a higher dimensionality than  six   if and only if  the function $D(\Omega) = \Omega^k \equiv  |\nabla u|^{2k}$, $k \neq 0$  (up to the equivalence transformations from the group $\mathcal{E}$). The  relevant MAI   is  seven-dimensional  and  generated by the operators \eqref{4} and the operator
      \begin{equation}\label{5}
        X_7 = 2kt \partial_t - u \partial_u.
    \end{equation}
\end{theorem}
\begin{corollary}
Equation  \eqref{0} with the diffusion coefficients $D$ of the form  \eqref{0*} is invariant with respect to the MAI
with the basic operators \eqref{4}.
\end{corollary}


\section{Reductions of the nonlinear equation  \eqref{0} to ODEs}

According  to Theorem 9,  equation  \eqref{0} possesses  its widest Lie symmetry in the case of the power-law  diffusivity.
Moreover, this case is important from the potential applicability point of view (see, e.g. \cite{barenb-1952}).
Thus, here we examine the nonlinear (1+2)-dimensional  equation
\begin{equation}\label{4-1}
    u_t = \text{div}\left(|\nabla u|^{2k} \nabla u \right).
\end{equation}

Using the seven-dimensional Lie algebra  with the basic operators \eqref{4}-\eqref{5} a wide range reductions of  \eqref{4-1} to  (1+1)-dimensional  PDEs and to  ODEs can be obtained. Here we restrict ourselves
 to an important case when \eqref{4-1} describes a process with radial symmetry w.r.t. to space variables (compare for instance an example in \cite{pascal-1992}). This allows to reduce  equation \eqref{4-1} to the form
\begin{equation}\label{4-2}
    U_t = \frac{1}{r} \left(r U_r^{2k+1} \right)_r
\end{equation}
(here we assume $U_r >0$, while the case $U_r <0$  leads to $  U_t = -\frac{1}{r} \left(r (-U_r)^{2k+1} \right)_r$, which can be treated in the same way) by the ansatz
\begin{equation}\label{4-3}
    u(t,x_1,x_2) = U(t,r), \  r=\sqrt{x_1^2+x_2^2}
\end{equation}
generated by the operator of  rotations $X_4 = x_2 \partial_{x_1} - x_1 \partial_{x_2}$.


First of  all we note that equation \eqref{4-2} with $k=-1$, i.e.
\begin{equation}\label{4-4}
    U_t = \frac{1}{r} \left(r U_r^{-1} \right)_r
\end{equation}
is transformed to the linear equation
\begin{equation}\label{4-5}
    V_t = -V_{zz}
\end{equation}
by the hodograph type substitution
\begin{equation}\label{4-6}
    r=\sqrt{V}, \quad  U=z.
\end{equation}
Thus, we assume in what follows that $k\not=-1$.  Moreover, we assume that $k\not=0$ (otherwise the linear radially-symmetric  heat equation is obtained) and   $k\not=- \frac{1}{2}$ because equation \eqref{4-2} degenerates to a linear ODE in this case.

 \noindent \emph{\textbf{Remark 6.}} In order to describe a diffusion process  equation \eqref{4-2} should be taken with the  sign of $2k+1$  on the  RHS.  Especially taking $k=-1$  one arrives at the linear heat equation instead of \eqref{4-5}. However, this observation does not affect any results presented below because both equations are equivalent up to the discrete transformation $t \to -t$.

 It can be easily calculated that  the MAI of \eqref{4-2} is four-dimensional Lie algebra with the basic operators
 \begin{equation}\label{4-7}
  X_0 = \partial_{U}, \  X_1 = \partial_t, \  D_0 =2(k+1)t\partial_t+r \partial_{r}, \  D_1 =kr \partial_{r}+ (k+1)U\partial_U
  \end{equation}
  i.e. the reduction of  \eqref{4-1} to the radial form partly inherits the relevant symmetries
 of this equation  (see \eqref{4} and \eqref{5})  and no others.
  A simple analysis shows that this Lie algebra leads   to four   essential different reductions of \eqref{4-2} to ODEs (any other reduction  is a composition of one of these four and the relevant group of transformations generated by Lie algebra \eqref{4-7}).  Four essentially different  combinations of  the operators from  \eqref{4-7}, the  relevant ans\"atze and ODEs are as follows.

  Case {\it (i)} $ X_1, \   U=\phi(r)$. ODE  for     $ \phi(r):$
   \begin{equation}\label{4-8}
   \left(r \phi_r^{2k+1} \right)_r=0.
   \ee

   Case {\it (ii)} $ D_0+\lambda X_0, \   U=\frac{\lambda}{2(k+1)}\log t +\phi(\omega), \  \omega=r t^{-\frac{1}{2(k+1)}}$. ODE for    $ \phi(\omega):$
   \begin{equation}\label{4-9}
   (2k+1)\phi^{''} +\omega^{-1}\phi^{'} +\frac{1}{2(k+1)}(\phi^{'})^{1-2k} - \frac{\lambda}{2(k+1)}(\phi^{'})^{-2k}=0.
   \ee

    Case {\it (iii)} $ D_1+\lambda X_1, \   U=e^{\frac{(k+1)t}{\lambda}}\phi(\omega), \  \omega=r e^{-\frac{kt}{\lambda}}, \ \lambda\not =0$. ODE  for     $ \phi(\omega):$
   \begin{equation}\label{4-10}
   (2k+1)\phi^{''} +\omega^{-1}\phi^{'} +\frac{k}{\lambda}\omega(\phi^{'})^{1-2k} - \frac{ k+1}{\lambda}\phi(\phi^{'})^{-2k}=0.
   \ee
   The subcase with  $\lambda =0$ leads to the separable ansatz $ U=r^{\frac{k+1}{k}}\phi(t)$, which produces the first-order ODE
    \begin{equation}\label{4-11}
   \phi^{'}=\Bigl(\frac{3k+1}{k}\Bigr)\Bigl(\frac{k+1}{k}\Bigr)^{1+2k}\phi^{1+2k}.
   \ee

   Case {\it  (iv)} $ D_0+\lambda D_1, \   \lambda\not =0, \  U= t^{ \frac{\lambda}{2}}\phi(\omega), \  \omega=r t^{-\gamma}, \ \gamma=\frac{1+\lambda k}{2(k+1)}$. ODE for     $ \phi(\omega):$
   \begin{equation}\label{4-12}
   (2k+1)\phi^{''} +\omega^{-1}\phi^{'} +\gamma \omega(\phi^{'})^{1-2k} - \frac{\lambda}{2}\phi (\phi^{'})^{-2k}=0.
   \ee

   It is a simple task  to integrate the nonlinear ODEs  \eqref{4-8} and  \eqref{4-11} because of their simplicity  and we omit this.
   The nonlinear ODEs  \eqref{4-9}, \eqref{4-10}  and  \eqref{4-12} are  not integrable
   for arbitrary  parameters (i.e. similarity exponents).

    ODE  \eqref{4-9}  is equivalent to the first-order ODE
   \begin{equation}\label{4-13}
   (2k+1)y^{'} +\omega^{-1}y +\frac{1}{2(k+1)}y^{1-2k} - \frac{\lambda}{2(k+1)}y^{-2k}=0,
   \ee
    where $ y(\omega)=\phi^{'}$. To the best of our knowledge  the general solution of  ODE  \eqref{4-13} for arbitrary parameters  $k$ and $\lambda $ is unknown (at least in terms of elementary functions). However, we note that
   this equation for $\lambda=0$ is a Bernoulli ODE and  can hence be linearised.
   Hence   the   general solution of ODE  \eqref{4-9} with  $\lambda=0$ of the form
   \begin{equation}\label{4-14}
\phi=\int \Bigl( \frac{-k\omega^2+C_1\omega^{-2k/(2k+1)}}{2(k+1)(3k+1)}\Bigr)^{\frac{1}{2k}}d\omega+C_2 \ee
is constructed  (hereafter $C_1$ and $ C_2$ are arbitrary constants).
Thus,  one obtains a two-parameter family of exact solutions for the nonlinear (1+2)-dimensional  equation  \eqref{4-1} of the form
  \begin{equation}\label{4-15}
u=\int \Bigl( \frac{-k\omega^2+C_1\omega^{-2k/(2k+1)}}{2(k+1)(3k+1)}\Bigr)^{\frac{1}{2k}}d\omega+C_2, \quad  \omega=t^{-\frac{1}{2(k+1)}}\sqrt{x_1^2+x_2^2}. \ee
Obviously, the integral in the RHS of   \eqref{4-15}  can be in terms of elementary functions
in some  cases, for example,   for  $C_1=0$  and for any $C_1\not=0$  provided $k=\frac{1}{2}, \frac{1}{4}, \frac{1}{6}\dots.$

We next consider  ODE  \eqref{4-10}, which, unlike  to the previous case,   cannot be reduced to a first-order ODE for arbitrary parameters  $k$ and $\lambda $. However, this equation can be rewritten in the form
\[ \left(\omega (\phi^{'})^{1+2k} \right)^{'}= \frac{ k+1}{\lambda}\omega \phi - \frac{ k}{\lambda}\omega^2 \phi^{'}  \]
and one may note the special case $k+1=-2k$, i.e. $k= -\frac13 $, when this equation is  equivalent to the first-order ODE
\[ \omega (\phi^{'})^{\frac13}= \frac{ 1}{3\lambda}\omega^2 \phi +C_1. \]
This ODE  with  $C_1\not=0$  has no solutions in terms of elementary functions while  setting $C_1=0$   one easily finds
 \begin{equation}\label{4-16}
\phi= \pm \frac{2}{\sqrt{C_2- 2(3\lambda)^{-3}\omega^4}}. \ee
Thus,  one obtains  exact solutions of the nonlinear (1+2)-dimensional  equation  \eqref{4-1} with  $k=-{\frac13}$ of the form
 \begin{equation}\label{4-17}
u= \pm \frac{2}{\sqrt{C_2e^{-\frac{2t}{3\lambda}}- 2(3\lambda)^{-3}(x_1^2+x_2^2)^2}}. \ee

Finally we turn to the nonlinear  ODE \eqref{4-12}.
A detailed analysis of this equation was undertaken   in \cite{barenb-1952} (see equation (3.6) with $k=1$ therein). In particular, it was noted that there is special case  $\gamma=-\frac{\lambda }{4}$, i.e., $\lambda=-\frac{2 }{3k+1}$ (see \eqref{4-12} above) when this equation can be reduced to the  first-order ODE
 \[ \omega (\phi^{'})^{2k+1}= \frac{ 1}{2(3k+1)}\omega^2 \phi +C_1 \]
 in the same   way as we have done  above for ODE  \eqref{4-10}. The exact solution obtained (see formulae (2.3)   in \cite{barenb-1952}) was  used for solving the Cauchy problem with the  Dirac function as the  initial profile.

\section{Discussion}

  In this paper,   complete descriptions of Lie symmetries of equation  (\ref{0***})
  in 1-D and 2-D  spaces  (i.e. one and two space variables, respectively)
have been obtained.  As one may note, the results depend essentially on the space dimensionality of
 (\ref{0***}) (so that  3-D and higher-dimensionality cases  should be also examined) and on the form of  the pair $(D, Q)$. We have established a special case (see Theorem 1) when this equation  in  1-D  space possesses an  infinite-dimensional Lie algebra of invariance so that it  is linearisable.  We also have found the special diffusivity $D=|\nabla u|^{-2}$, which is special one of   (\ref{0***})  in  2-D  space (see Table 3).
 It should be stressed that   equation (\ref{0***})  depending on    the form of $(D, Q)$   admits a surprisingly
rich  set of Lie symmetry algebras (see Tables 1 and 2).  Several of these Lie symmetries  would be unlikely
to be identified by a simple extrapolation from  the known results \cite{dor, dor-svi} for the standard reaction-diffusion
equation
\[   u_t = \text{div}\left( D(u) \nabla u \right) +Q(u). \]

We have examined separately equation \eqref{0} with $n=2$, which sometimes is called the Perona-Malik equation,  because of  its  importance for direct applications.
  In contrast to the (1+2)-dimensional  nonlinear diffusion equation \eqref{1a}, i.e.
\[ u_t = \text{div}\left( D( u) \nabla u \right), \]
  which admits a wide range  of MAIs depending on the form  the diffusivity  $ D( u)$
\cite{narib1970},  this  alternative equation     admits  a unique extension of the MAI (see Theorem 9).
In this   sense equation   \eqref{0}   is not  analogous to   the standard nonlinear diffusion equation \eqref{1a}
 (at least from the Lie symmetry point of view). Moreover,  equation \eqref{0} with the common
nonlinearities  \eqref{0*}  does not lead to any extension of  the Lie symmetry.
 We believe it plausible  that a search for conditional symmetries of the nonlinear (1+2)-dimensional equation \eqref{0} in order to compare  with the known result for the (1+2)-dimensional  nonlinear diffusion equation \eqref{1a}\cite{arr-goard-br} could lead to more optimistic results.

On the other hand, one may  consider  the (1+2)-dimensional equation
\begin{equation}\label{5-2}
    u_t =  D( \Omega) \nabla^2 u ,
\end{equation}
instead of   \eqref{0}. In contrast to the (1+1)-dimensional case, equation  \eqref{5-2} is not
equivalent to   \eqref{0}.
On the other hand, this equation possesses an  infinite-dimensional Lie algebra in the case  $D( \Omega)=\Omega^{-1} $. Moreover this algebra
contains
 the operators of the form
\[ X^{\infty} = A(x_1,x_2) \partial_{x_1} + B(x_1,x_2) \partial_{x_2}  \]
where the pair $(A(x_1,x_2),B(x_1,x_2))$ is an arbitrary solution of the Cauchy-Riemann system,
having similar structure to the operators
\[ X^{\infty} = A(x_1,x_2) \partial_{x_1} + B(x_1,x_2) \partial_{x_2} - 2 A_{x_1} u \partial_{u} \]
that  arise  in the MAI for the nonlinear diffusion equation \eqref{1a} with   $D (u)=u^{-1}$.
Thus, investigation of Lie symmetry properties of equation \eqref{5-2} in higher dimensions
may  lead to essentially  different results comparing with  \eqref{0}.

Finally, we have  reduced   the (1+2)-dimensional  equation \eqref{0} with a  power-law diffusivity, i.e. equation   \eqref{4-1},  to a corresponding
  (1+1)-dimensional equation in order to find exact solutions. It turns out that the equation obtained still possesses a  rich symmetry,  hence  further reductions to ODEs have been performed. As a  result,    two families of new  exact solutions of  equation \eqref{4-1} have been constructed (see formulae  \eqref{4-15} and  \eqref{4-17}) and  solutions found much earlier in \cite{barenb-1952}
have been recovered.

\section{Acknowledgements}
This research was supported by a Marie Curie International Incoming
Fellowship to the first author within the 7th European Community
Framework Programme.

\end{document}